\documentclass[superscriptaddress,twocolumn]{revtex4}

\usepackage{graphicx,color,url,ulem}
\usepackage{amssymb}

\definecolor{darkgreen}{RGB}{0,139,0}
\definecolor{turqoise}{RGB}{64,224,208}
\def\myeightstar{\protect\makebox[0pt][l]{+}$\times$}
\definecolor{b}{rgb}{0,0,1.0}
\definecolor{r}{rgb}{1,0,0}
\definecolor{g}{rgb}{0,1,0}

\def\T{\textit{\textbf{T}}}
\def\thetaav{\theta_\mathrm{av}}

\begin{document}

\newcommand{\SZFKI}{Institute for Solid State Physics and Optics, Wigner Research Center for Physics,
Hungarian Academy of Sciences, P.O. Box 49, H-1525 Budapest, Hungary}

\newcommand{\DUI}{Faculty of Physics, University of Duisburg-Essen, 47048 Duisburg, Germany}

\newcommand{\WAR}{Department of Physics and Centre for Complexity Science, University of Warwick,
Coventry CV4 7AL, UK}

\newcommand{\MDB}{Otto-von-Guericke-University, D-39106 Magdeburg, Germany}

\title{Orientational order and alignment of elongated particles induced by shear}

\author{Tam\'as B\"orzs\"onyi}

\email{borzsonyi.tamas@wigner.mta.hu}
\affiliation{\SZFKI}
\author{Bal\'azs Szab\'o}
\affiliation{\SZFKI}
\author{G\'abor T\"or\"os}
\affiliation{\SZFKI}
\author{Sandra Wegner}
\affiliation{\MDB}
\author{J\'anos T\"or\"ok}
\affiliation{\DUI}
\author{Ell\'ak Somfai}
\affiliation{\WAR}
\author{Tomasz Bien}
\affiliation{\MDB}
\author{Ralf Stannarius}
\affiliation{\MDB}

\begin{abstract}
Shear induced alignment of elongated particles is studied experimentally and numerically.
We show that shear alignment of ensembles of macroscopic particles is comparable even on a quantitative level to simple
molecular systems, despite the completely different types of particle interactions. We demonstrate
that for dry elongated grains the preferred orientation forms a small
angle with the streamlines, independent of shear rate across three decades.  
For a given particle shape, this angle decreases with increasing aspect ratio of the particles. The shear-induced alignment results in a considerable reduction of the effective friction of the granular material.
\end{abstract}

\maketitle

Flow of large ensembles of elongated objects -- often observed in nature or industry --
usually induces pronounced alignment of the building blocks. This phenomenon is found
at all length scales, in log jams on rivers, in slowly deformed multicomponent clastic rocks,
in seeds, nanorods, viruses, and even at molecular scales in nematic liquid crystals
\cite{mach2009,or1995,kuna2006,fabr2005,ledo2001,ledo2005,brzh2011,beba1994,ehch2003,pega2007,ca2011,skla1979,ga1972,beje1985,ehhe1995,haho2004,kasz2002}.
On one hand, such alignment processes are poorly characterized for macroscopic objects,
even though granular flows have been extensively studied in the last two decades
\cite{beba1994,ehch2003,azra2010,fehe2003,feme2004,lu2008,brto2011,pega2007,ca2011,jana1996,arts2006,jofo2006,gdrmidi2004,lobo2000,crem2005,boec2009}.
On the other hand, shear alignment and collective reorientation dynamics is well documented and
exploited at molecular scale (for nematic liquid crystals) and can be
described quantitatively by continuum theory \cite{le1979,je1978}.
Studies have also shown that the stationary alignment observed for simple
one component systems may be replaced by a so called {\it tumbling} behavior for more complex
materials, e.g. nematic liquid crystals with pre-transitional smectic fluctuations
\cite{beje1985,je1978,mape1995,sala2009}, polymeric liquid crystals
\cite{cabu2001,bubr2005,siqu2007,falu1994} as well
as dilute suspensions where individual particles rotate in shear flows
(Jeffery orbits, \cite{je1922}). For the limiting
case of very dense suspensions direct interparticle forces dominate,
leading to stationary flow alignment, analogously to what is expected for simple one component systems.

In this study we focus on simple one component systems and aim to bridge the gap
between systems of very different sizes and interaction types.
Our goal is to determine the shear induced order and average orientational
angle for macroscopic objects, to track how the phenomenon depends on the particle aspect
ratio, and find to what extent the applied shear rate is relevant.
Starting from an initially random orientation, we also aim to quantify the change
in the shear resistance of a sample due to the evolving alignment.

\begin{figure}[htb]
\includegraphics[width=\columnwidth]{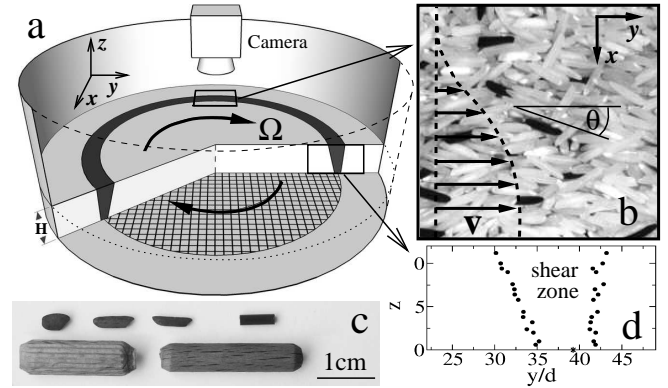}
\caption{(a) Schematic illustration of the split bottom shear cell \cite{fehe2003}.
The granular material consisting of elongated grains with mean
diameter $d$ and length $L$ is sheared due to the rotation of the
thin circular plate below the material with diameter $D_p\approx 80\, d$
and angular velocity $\Omega$.
The container diameter is $D_c\approx 140\,d$, while typical filling height
is $H\approx11d\approx0.14\,D_p$ (unless stated otherwise). Shear deformation is
localized at the region marked with dark gray. 
(b) Particles are oriented on average 
at a small angle $\theta$ with respect to the streamlines.
(c) Photographs of the particles studied, with $L/d= 2.0$, $3.4$, $4.5$ (rice),
as well as $3.5$ (glass cylinders) and $3.3$, $5.0$ (pegs).
(d) Measured borders of the shear
zone for pegs with $L/d= 5.0$ (border defined at approximately 10$\%$ of the maximum shear rate).
}
\label{setup}
\end{figure}

We sheared the granular material in a cylindrical split bottom geometry
(see Fig.~\ref{setup}) where stationary flow is obtained by a continuous
rotation of a thin plate below the material.
The strain is localized
\cite{fehe2003,feme2004,diwa2010} in the so called shear zone (indicated with dark gray).
In this geometry the time and ensemble averaged shear rate changes with the radius in a relatively
wide region allowing us
to map the shear rate dependence of the alignment in a single experiment \cite{lu2008}.
Prolate particles with different
length to diameter ratios $L/d$ were investigated (Fig.~\ref{setup}c).
Two methods were used to visualize and evaluate shear-induced alignment:
(i) complete 3D imaging of the sample with X-ray tomography and
(ii) optical imaging of the particles at the upper surface of the system with a digital camera.
For the characterization of changes in effective friction due to the
alignment, the torque on the
bottom plate was measured at stationary rotation.
The experiments were complemented with numerical simulations using the
Discrete Element Method (DEM). In the simulations elongated particles
were prepared by numerous overlapping spheres and the alignment of the
particles was measured in plane Couette flow.

The shear-induced orientational order is monitored by diagonalizing the
symmetric traceless order tensor \T{}:
\begin{equation}
T_{ij}= \frac{3}{2N} \sum\limits_{n=1}^N \left[{\ell}^{(n)}_i {\ell}^{(n)}_j -\frac{1}{3} \delta_{ij}
\right] \quad ,
\end{equation}
where $\vec {\ell}^{(n)}$ is a unit vector along the long axis of particle $n$, and the sum is over
all $N$ detected particles. The largest eigenvalue of \T{} is taken as the primary order parameter $S$.
A second, biaxial order parameter $D$ is defined as the difference of the two other eigenvalues of \T.
The shear-induced alignment is characterized by the average alignment angle $\thetaav$
measured with respect to the shear flow direction (see Fig.~\ref{setup}b), i.e. the deviation 
of the largest main axis of \T{} from the streamlines.

\begin{figure}[ht]
\includegraphics[width=\columnwidth]{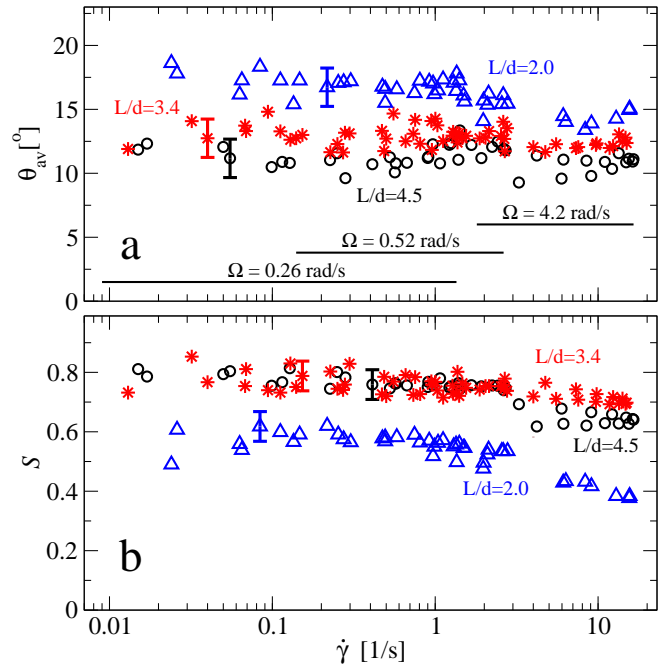}
\caption{(color online). (a) The average orientational angle $\theta_{\rm
av}$ of the particles with respect to the streamlines and (b) the order
parameter $S$ as a function of the shear rate $\dot{\gamma}$. Both panels
correspond to optical measurements for rice grains with length to width
ratios $L/d= 2.0$, $3.4$, and $4.5$. The value of $\theta_{\rm av}$
is roughly constant, and $S$ varies only very weakly across three decades
of the shear rate.
}
  \label{theta-gamma}
\end{figure}

First, we present the shear rate dependence of the average alignment angle $\thetaav$
and the order parameter $S$ obtained by digital optical imaging at the surface layer.
In this experiment, the shearzone at the top was divided into 15 bands
(widths 0.19 cm $\approx d$) and the average shear rate was
determined for each band.
The orientation and in-plane length of colored tracer particles (6$\%$ of the sample)
were detected with a particle tracking code, yielding about 60,000 particle positions
in each band.
The center $\theta_{\rm av}$ of the orientational distribution was determined by fitting a
Gaussian, and the order parameter $S$ was calculated.
These two quantities are shown for the roughly ellipsoidal rice grains
in Figs.~\ref{theta-gamma}a,b as a function of the shear rate $\dot{\gamma}$.
Here the shear rate spans
a wide range of inertial numbers, about 0.0002 to 0.4, using 3 different
rotation rates of the bottom plate ($\Omega=$ 0.26, 0.52 and 4.2 rad/s) and 
considering the spatial variation of $\dot{\gamma}$ in
the shear zone. The ranges of $\dot{\gamma}$ spanned for each $\Omega$
are indicated with horizontal bars in Fig.~\ref{theta-gamma}a. 
(The inertial number
\cite{gdrmidi2004} is defined as $I=\dot{\gamma}d\sqrt{\rho/P}$, where
$\rho$ is the density of particles and $P$ is the pressure.)

As it is seen, $\theta_{\rm av}$ is well defined,
and it is essentially independent of the shear rate for the
the whole range of $\dot{\gamma}$ investigated here. This result is
not unexpected, it is actually similar to the case of flow aligning nematic
liquid crystals consisting of elongated molecules, where the alignment angle
is a material parameter, i.e. shear rate independent as well.
But even if the aspect ratios of the building blocks (molecules and granulates) are similar,
the interparticle forces have very different character in these two systems.
While the molecules experience attractive forces (dipole-dipole, van der Waals)
the grains investigated here interact only via hard core repulsion
and friction. The fact that despite of these differences the alignment angle
is independent of the shear rate underlines the robustness and the
geometrical origin of the shear alignment.
The slightly decreasing value of the order parameter $S$ with
increasing shear rate (Fig.~\ref{theta-gamma}b) shows that
the inertial effects make the material less ordered while preserving the
same alignment direction.

The numerical simulations were performed using the discrete element method (DEM),
where long particles were approximated by chains of overlapping spheres.
Special care was taken to reduce the possible
interlocking of neighboring chains: both the overlap (30-70\%) and the
number of participating spheres were varied randomly.  For implementing
a soft particle model we used the general purpose molecular dynamics
code LAMMPS \cite{pl1995}, extended with particle cementing interactions within
the chains \cite{brto2011} without allowing the contacts to break. An assembly of
600-1200 elongated particles was confined under constant
pressure between two parallel, rough walls of finite mass; the system
was periodic in the other two directions. The shear flow was realized
by imposing a fixed, opposite velocity of the walls.
In the steady state the shear velocity profiles were reasonably linear to neglect
shear rate variations.  The inertial number was in the range 0.004 to 0.09.
For calculating the average alignment angle and order parameter only the central
60\% was considered, in order to exclude influences of the walls.

The complete 3D arrangement of all particles in the shear zone was also determined
experimentally for two samples (pegs with $L/d = 3.3$ and $5.0$) using X-ray tomography \cite{x-ray}.
The purpose of these experiments is to show that the order and alignment data extracted from
surface particles are representative for the global order in the shear zone, with only slight variations.
The particles were detected by applying a watershed algorithm to the recorded images.
The components of \T{} were determined for a surface region of $\sim 1$~cm
(two particle diameters) depth and for a 2~cm thick layer below (bulk), corresponding to about
10,000 and 20,000 particle positions, respectively.

\begin{figure}[ht]
\includegraphics[width=\columnwidth]{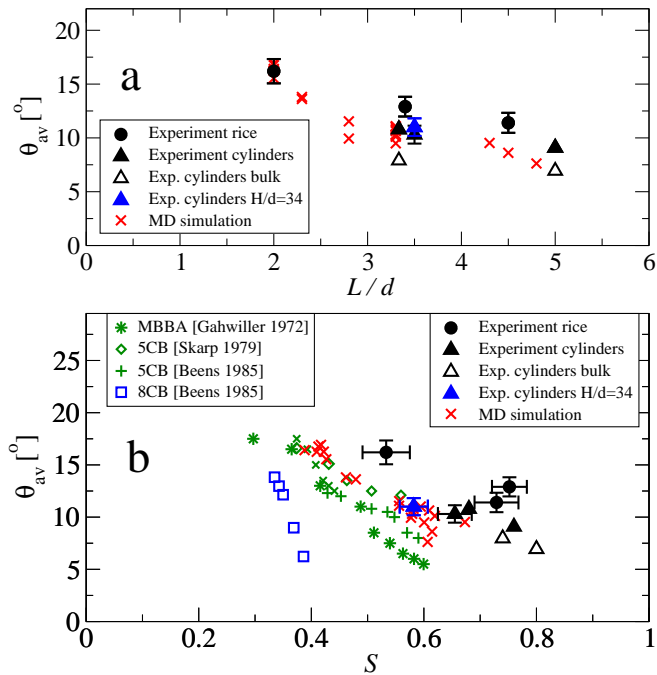}
\caption{(color online). The average orientational angle $\theta_{\rm av}$
as a function of the (a) length to diameter ratio $L/d$ of the particles
and (b) order parameter of the system obtained by experiments 
($\bullet$,$\blacktriangle$,$\vartriangle$)
and numerical simulations ({\color{red}$\times$}). One datum point
({\color{blue}$\blacktriangle$}) was taken with a particularly deep layer, $H/d=34$.
The alignment angles for typical flow aligning nematic liquid crystals MBBA and 5CB
({\color{darkgreen}\myeightstar,{\large$\diamond$},+})
\cite{skla1979,ga1972,beje1985,chch2004,ehhe1995}
and for a nematic liquid crystal 8CB ({\color{blue} $\Box$}) which shows tumbling when
approaching the smectic A phase transition are shown for comparison.
 }
  \label{theta-s}
\end{figure}

Figure \ref{theta-s} combines the bulk and surface results from X-ray tomography,
 the surface data from optical experiments, and the numerical results.
Comparing the average values of the shear alignment angle for the six samples
investigated, we find a systematic decrease of $\theta_{\rm av}$ with increasing
length to width ratio (Fig.~\ref{theta-s}a) for a given material (rice or cylinders).
At comparable aspect ratios, the alignment angle
appears to be smaller for cylinders (as if they had a larger effective aspect ratio).
Fig.~\ref{theta-s}b shows that the alignment angle $\theta_{\rm av}$ displays
a decreasing trend with increasing order parameter both for experiment and simulation.
The measurement with $H/d=34$ shows that the value of $\theta_{\rm av}$ is
independent of $H$. The orientational order slightly increases with decreasing $H$, this
may be an interesting subject for further investigations.
As a possible explanation for the trends observed in Figs.~\ref{theta-s}a,b,
we suggest the following picture: in a dense granular system,
the elongated particles are entangled with neighbors that hinder their rotation.
In a shear flow an orientation distribution develops due to the competition
between the agitations by collisions and this hindering effect.
The orientation angles of the particles have a distribution centered around
$\theta_{\rm av}$, where the width of the distribution roughly equals $\theta_{\rm av}$.
For more elongated and therefore
more entangled particles the angular fluctuations will be smaller, so the width of
the distribution is smaller, making $\theta_{\rm av}$ smaller.

The X-ray tomograph measurements confirm that the shear alignment is very similar at the surface and
in the bulk, with slightly larger order parameter $S$ and slightly smaller alignment angle
$\thetaav$ in the bulk (see Figs.~\ref{theta-s}a and b) compared to the surface.
The slightly larger order observed in the bulk is connected to a smaller
standard deviation from the mean orientation within the shear plane (while the out-of-plane
standard deviation is similar in the bulk and at the surface). This is in good agreement
with molecular dynamics simulations of a hard-spheroid fluid under shear flow \cite{yual1997},
where a similar biaxiality was detected.
For our case the measured biaxiality is somewhat stronger in the bulk ($D \approx 0.09$)
than at the surface ($D\approx 0.03$).

For comparison, we have also included the order parameter dependence of the
alignment angle for typical flow aligning nematic liquid crystals (MBBA and 5CB)
for which the aspect ratio of the molecular shape is estimated to be $L/d=2.5$ \cite{ehhe1995}.
For molecular nematics the order parameter is independent of the shear rate (in this case determined
by the temperature).
We emphasize that the decreasing tendency of the $\theta_{\rm av}(S)$ curves is very similar
for granulates and for these nematics. We also note that different flow behavior is observed for some
nematics that exhibit strong smectic fluctuations
when approaching the smectic A phase transition \cite{beje1985,je1978,mape1995,sala2009}.
For example, nematic
8CB shear aligns only in a narrow temperature range near the clearing point. With decreasing temperature,
the $\theta_{\rm av}$ curve approaches zero when one of the viscosity coefficients changes sign, 5.6 K
above the transition to smectic A
(where $S \approx 0.4$). Pretransitional smectic clusters drive the system into a tumbling behaviour, where
the director continuously rotates in the shear flow.
Another example for tumbling is the case of polymeric liquid crystals, where the presence of
polymeric chains strongly modifies the flow behavior \cite{cabu2001,bubr2005,siqu2007,falu1994}.
For molecular systems the flow behavior of the aligned case as well as the dynamical modes of
the tumbling state can be accurately characterized by modern experimental tools, such as X-ray
scattering \cite{cabu2001,bubr2005,siqu2007}, NMR \cite{falu1994,siqu2007} or conoscopy \cite{mape1995},
and non-equilibrium molecular dynamics simulations \cite{yual1997,sala2009}.
Various theories have also predicted the instabilities and different dynamical modes, like tumbling, kayaking,
wagging, log rolling or flow aligning \cite{kudo1984,la1990,fowa2003}. These theories are more general in
the sense that they map the phase diagram not only as a function of aspect ratio but also as a function
of density including the case of dilute systems. They predict flow alignment for our parameter
range \cite{kudo1984,la1990}.
Altogether, the fact that in our dense granular system, with no detectable smectic fluctuations and
simple interactions between the particles stationary alignment is observed, is in agreement with the
above observations and theories.

The initial evolution of order has been measured starting from
\begin{figure}[ht]
\includegraphics[width=\columnwidth]{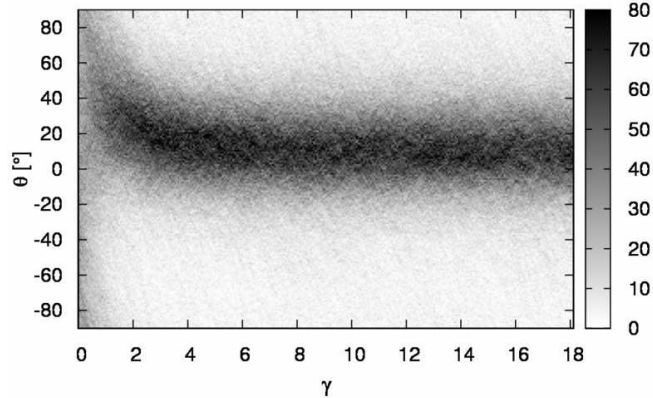}
\caption{
Evolution of the orientational distribution of the particles starting
from an initially random orientation.
}
  \label{transient}
\end{figure}
a randomly oriented configuration. Data were obtained from 1300 experiments.
The orientational distribution is presented in Fig.~\ref{transient}
as a function of the strain taken in the middle part of the shear zone for a
sample with $L/d=3.5$.
As it is seen, order develops quickly and the alignment angle $\theta_{\rm av}$ decreases
towards the stationary value. The strong decrease of $\theta_{\rm av}$ at the
beginning of the process is a consequence of the fact that the vast
majority of particles rotate in the same direction due to the shear flow.
The transient in Fig.~\ref{transient} evidences that an equilibrium alignment is
established after the sheared particles have passed approximately three to four neighbors.

\begin{figure}[ht]
\includegraphics[width=\columnwidth]{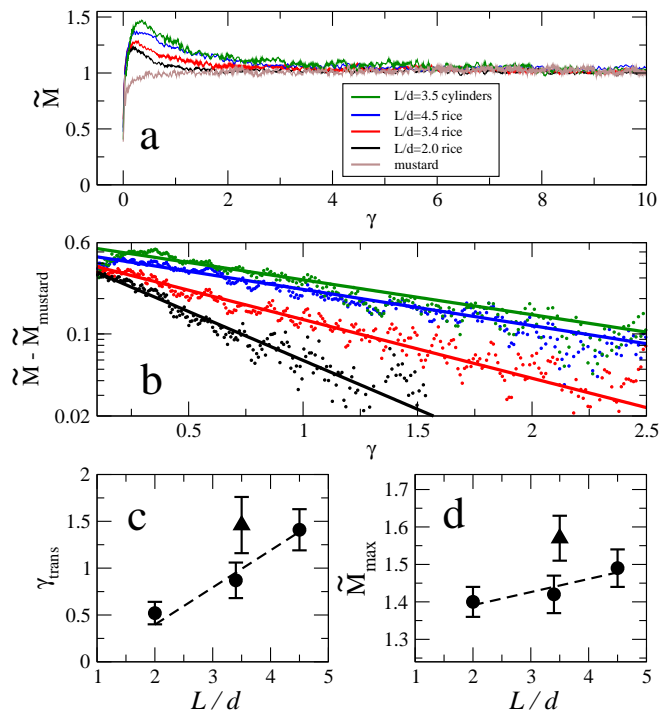}
\caption{
(color online). (a)-(b) Evolution of the applied torque needed to maintain stationary rotation
of the bottom plate starting from an initially random orientation. Panel (a) also includes a reference
dataset obtained for spherical particles which is then subtracted from the other
datasets resulting in the curves presented in panel (b). (c) Characteristic strain $\gamma_{\rm trans}$
corresponding to the transient and (d) ratio of the torques $M_{\rm max}/M_\infty=\widetilde{M}_{\rm max}$
needed to rotate an unoriented sample and the oriented sample (stationary state)
for rice ($\bullet$), and cylinders ($\blacktriangle$).
}
  \label{torque}
\end{figure}

The increasing orientational order and the development of the shear alignment
angle leads to a systematic change of the effective friction of the sample at
the beginning of the process. This can be quantified by measuring the torque
$M$ needed to rotate the circular bottom plate with a constant speed. As a
reference we have also recorded the torque for nearly spherical mustard seeds.
This helps us to correct for the transient effects which are not directly
related with orientational changes.
The signals were corrected by subtracting the torque needed for rotating the
empty plate and rescaled by the stationary value $M_\infty$.
The rescaled torque $\widetilde{M}=M/M_\infty$ is shown in Fig.~\ref{torque}a,
while Fig.~\ref{torque}b presents the curves for the elongated grains after
subtraction of the data for mustard seeds on a log-lin scale.
The evolution of the internal friction of the material can be characterized
by an exponential decay of the excess torque as
$\widetilde{M}  -\widetilde{M}_{\text{mustard}} \approx (\widetilde{M}_{\rm max}-1)
e^{-\gamma/\gamma_{\text{trans}}}$.
The solid lines in Fig.~\ref{torque}b are exponential fits, they define a
characteristic deformation $\gamma_{\text{trans}}$ describing the transient. 
The dependence of $\gamma_{\text{trans}}$ on the aspect ratio is shown in 
Fig.~\ref{torque}c. For rice grains, the transients increase with grain length. 
This evidences a slower dynamics of the more elongated particles.
A linear fit yields $\gamma_{\text{trans}} =  0.4 (L/d-1)$ for rice.
The ratio of the torques corresponding to the totally random and the oriented
systems is estimated around $\widetilde{M}_{\rm max} \approx 1.5$ as seen
in Fig.~\ref{torque}d.

In conclusion, we used surface optical imaging, bulk X-ray tomography, as
well as numerical simulations to study the shear flow of elongated dry
granular particles.  We found that the particles quickly orient themselves
around a preferred direction, which is at an angle from the streamlines.
This angle is almost independent of the shear rate (or inertial number)
across three decades, and for a given class of shapes (rice or cylinders)
it decreases with increasing particle aspect ratio.
This shear alignment is a robust phenomenon, and its geometric origin is
underlined by the similarity with nematic liquid crystals, despite the
completely different interparticle forces.  The
shear alignment decreases the effective friction by about one third when
compared to the initial randomly oriented state.

The authors are thankful for G. Rose, chair of the department for
Healthcare Telematics and Medical Engineering of the
Otto von Guericke University, Magdeburg, and appreciate discussions with
H. Brand, K. Daniels, A. J\'akli, A. Krekhov and H. Pleiner.
J.T. acknowledges the support of the German Research Foundation (DFG
grant BR 3729/1). T.B. acknowledges Magdeburg University for a visiting grant.

\end{document}